# Ultrasensitive hybrid optical skin

Lei Zhang[1]*, Jing Pan[1], Zhang Zhang[1], Hao Wu[1], Ni Yao[1], Dawei Cai[1], Yingxin Xu[1], Jin Zhang[1], Guofei Sun[2], Liqiang Wang[1], Weidong Geng[2], Wenguang Jin[3], Wei Fang[1], Dawei Di[1,4] and Limin Tong[1]*

[1] State Key Laboratory of Modern Optical Instrumentation, College of Optical Science and Engineering, Zhejiang University, Hangzhou 310027, China.

[2] College of Computer Science and Technology, Zhejiang University, Hangzhou 310027, China.

[3] College of Information Science and Electronic Engineering, Zhejiang University, Hangzhou 310027, China.

[4] Cavendish Laboratory, University of Cambridge, JJ Thomson Avenue, Cambridge CB3 0HE, United Kingdom.

*Correspondence to: zhang_lei@zju.edu.cn; phytong@zju.edu.cn.

**Abstract:** Electronic skin, a class of wearable electronic sensors that mimic the functionalities of human skin, has made remarkable success in applications including health monitoring, human-machine interaction and electronic–biological interfaces. While electronic skin continues to achieve higher sensitivity and faster response, its ultimate performance is fundamentally limited by the nature of low-frequency AC currents in electronic circuitries. Here we demonstrate highly sensitive optical skin (O-skin) in which the primary sensory elements are optically driven. The simple construction of the sensors is achieved by embedding glass micro/nanofibers (MNFs) in thin layers of polydimethylsiloxane (PDMS). Enabled by the highly sensitive power-leakage response of the guided modes from the MNF upon external stimuli, our optical sensors show ultrahigh sensitivity (1870 kPa$^{-1}$), low detection limit (7 mPa) and fast response (10 μs) for pressure sensing, significantly exceeding the performance metrics of state-of-the-art electronic skins. Electromagnetic interference (EMI)-free detection of high-frequency vibrations, wrist pulse and human voice are realized. Moreover, a five-sensor optical data glove and a 2×2-MNF tactile sensor are demonstrated. Our results pave the way toward wearable optical devices ranging from ultrasensitive flexible sensors to optical skins.

## 1. Introduction

Over the past decades, breakthroughs in flexible electronics and nanotechnology have enabled wearable electronic sensors[1,2] that respond to external stimuli via capacitive[3], resistive[4], piezoelectric[5] and triboelectric[6] effects. While microelectronic technologies[7-12] continue to push wearable devices towards higher sensitivity, faster response, better biocompatibility and denser integration, it may ultimately reach the limit imposed by the nature of low-frequency electromagnetic fields (i.e., AC currents). For example, response time is limited by parasitic effects and crosstalk in high-density electronic circuitries. In contrast, using photons instead of electrons as the signal carrier is an ideal strategy to circumvent these limitations[13], as has been shown in optical fiber sensors[14]. Conventional optical fibers, which are ~125 μm in diameter, are usually too thick and rigid to be made into a wearable device. Optical MNFs, which are about 1 μm in diameter, are capable of guiding light with high flexibility[15]. Owing to their atomically-precise geometric uniformities, they offer waveguiding losses (e.g., < 0.05 dB/cm)[16] much lower than any other optical waveguides of similar sizes, and mechanical strength (e.g., tensile strength > 5 Gpa[17]) higher than spider silks (0.5-3.6 GPa[18]). Recently, MNFs have been attracting increasing interest for optical sensing applications[19], but their potential for wearable devices is yet to be explored.

## 2. Results and discussion
### 2.1 Fabrication and characterization of MNF-embedded PDMS patches

In this work, highly uniform MNFs (Fig. 1a) are fabricated by taper drawing silica glass optical fibers, which are flexible in routing light at micrometer scale (Fig. 1b). However, the as-fabricated MNFs, with large fractional evanescent fields exposed to open air[20], are highly sensitive to environmental disturbance (e.g., direct physical pressure) or contamination (e.g., dust adsorption), which may lead to unpredictable variations of guided signals. To employ MNFs for wearable sensor applications, the guided optical fields must be well managed. Herein, we use a thin layer of PDMS, a highly flexible and biocompatible polymer with refractive index (n = 1.40) slightly lower than that of silica (n = 1.46), to enclose the MNF and isolate the evanescent fields, while maintaining high mechanical flexibility and low optical losses of the MNF. The layer thickness, from 1 mm (e.g., 0.8 mm in Fig. 1c) to less than 100 μm (e.g., 80 μm in Fig. 1d), is sufficiently thick to contain the evanescent fields of the MNF typically extending several microns. Figure 1c, d show typical MNF-embedded PDMS patches being bent in free space (Fig. 1c) or attached on human hand (Fig. 1d). Within a PDMS host film, the MNF can be made into various shapes (Fig. 1e), offering additional flexibility for stretching or bending. Meanwhile, at optical frequencies, the crosstalk between closely patterned MNFs can be minimized. Figure 1f shows the arrangements of two MNFs with 3-μm spacing (top panel) and perpendicular crossing with direct contact (bottom panel), respectively. With green or red optical signals waveguided in individual MNFs, no crosstalk is observed.

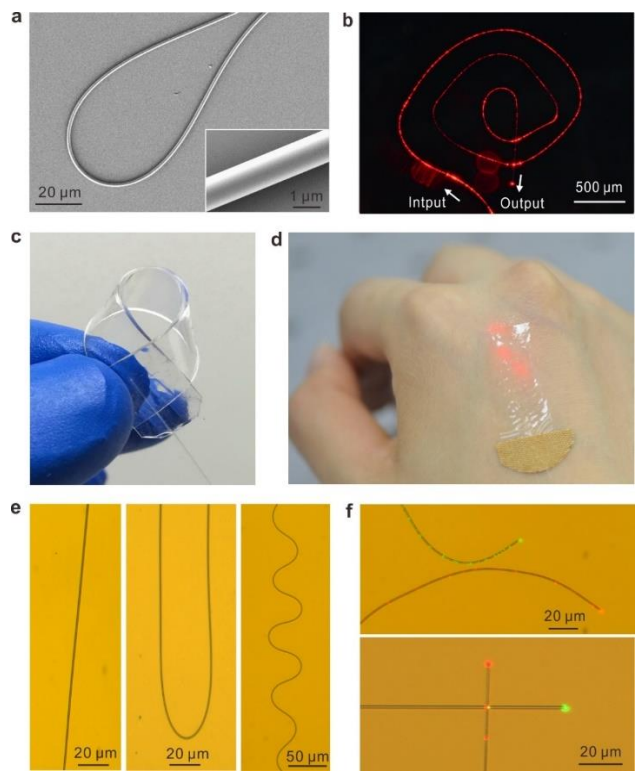

**Figure 1. Fabrication and characterization of MNF-embedded PDMS patches. a,** SEM image of a 900-nm-diameter glass MNF with a bending radius of 30 μm. Inset: close-up image of the MNF showing smooth surface and uniform diameter. **b,** Optical microscope image of a 1-μm-diameter glass MNF spiral guiding a 633-nm optical signal on a $MgF_2$ substrate. **c,** Photograph of a bent MNF-embedded PDMS patch. **d,** Photograph of a MNF-embedded PDMS patch attached on human hand. **e,** Optical microscope images of three patches with straight (left), bent (middle), and wavy (right) MNFs, respectively. **f,** Optical microscope images of two MNFs guiding 532- and 633-nm signals separately. The two MNFs are assembled into side-by-side (top) and perpendicular crossing structures (bottom), with no crosstalk observed.

## 2.2 Pressure response of an MNF-embedded PDMS patch

The pressure response of an MNF-embedded PDMS patch adhered on a glass slide (Fig. 2a) is investigated by measuring the optical transmission through the MNF. As shown in Fig. 2b, when the patch is slightly bent (e.g., at a bending angle of 5°, induced by pressure), the well confined symmetric mode of a 1-μm-diameter MNF at the input port evolves into an asymmetric profile with clear optical leakage after propagating merely 100 μm, making it an optical skin (O-skin) highly sensitive to micro-deformation. For example, upon a slight touch of a finger, light leakage from the O-skin is clearly observed (Fig. 2c). Figure 2d gives typical response of three 120-μm-thickness O-skins (with MNF diameters of 800 nm, 1.5 μm, and 2.8 μm, respectively) by laterally scanning a 10-kPa pressure perpendicular to the MNF. Using 790-nm probe light, the sensor starts to function from a lateral offset of about 700 μm (y = 700 μm), with maxima at zero offset (y = 0). Also, the thinner MNF gives higher sensitivity due to the larger fraction of evanescent fields. Meanwhile, using a tungsten lamp as the probe light, we are able to obtain broadband spectral response in a single measurement. Figure 2e shows broadband O-skin response to pressures from 2 to 10 kPa. With increasing wavelength, the output intensity of the MNF decreases and the sensitivity increases, as a result of the increasing fractional evanescent fields. The broadband wavelength-dependent response offers an opportunity for tuning the sensitivity and detection range in the same O-skin using different wavelengths, and thus broadens the dynamic range without sacrificing sensitivity in a single O-skin. In addition, by using purely optical effects, the O-skin is completely EMI-free. Figure 2f shows the response of an O-skin to finger actions of tapping, non-contact movement and dragging successively. The clear distinguishability of the three types of actions of the O-skin circumvents the EMI issue in electronic devices (e.g., capacity-sensitive sensors[21]) when being operated in a conductive medium. The long-term operational stability and excellent reliability of the O-skin is verified by alternately applying/removing a 2-kPa pressure for more than 10,000 cycles (Fig. 2g).

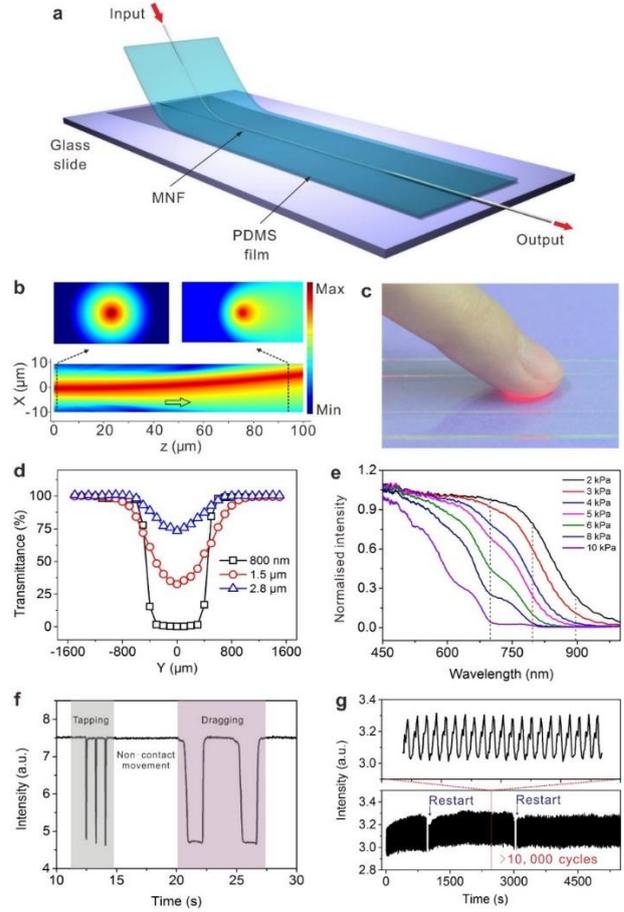

**Figure 2. Characterization of substrate supported O-skins. a,** Schematic of testing an O-skin on a glass slide. **b,** Optical field intensity distributions of 900-nm-wavelength light guiding along a 1-μm-diameter glass MNF embedded in a 5°-bent O-skin. **c,** Photograph of light leaking out of an O-skin upon a finger touch. **d,** Lateral pressure response of an O-skin with MNF diameters of 0.8, 1.5 and 2.8 μm, respectively. **e,** Pressure response of an O-skin with a wide spectral range. **f,** Typical response of an O-skin to finger movements of tapping, non-contact movement and dragging successively. **g,** Long-term operational characteristics of an O-skin measured by alternately applying/removing a 2-kPa pressure for more than 10,000 cycles.

## 2.3 Characterization of suspended and wearable O-skins

To function as a wearable sensor, the O-skin should conform to non-flat surfaces or be operational in a suspended mode. Figure 3a shows a schematic diagram of a suspended O-skin that responds to external force with a micro pit. Under the same pressure, a suspended O-skin allows higher degrees of deformation, offering better sensitivity than that attached on rigid surfaces (e.g., Fig. 2a). Figure 3b shows the response of a suspended O-skin (80-μm thickness, 980-nm MNF diameter), showing its ability to detect a water droplet down to 1 mg in weight that is too small to cause any observable deformation of the O-skin (inset of Fig. 3b). To examine the detection limit of this type of devices, a calibrated stream of airflow is used to apply a small pressure on a suspended O-skin (50-μm thickness, 780-nm-diameter MNF). The clear response of the O-skin to pressures down to 0.1 Pa with a high signal-to-noise ratio (Fig. 3c) corresponds to a detection limit of about 7 mPa.

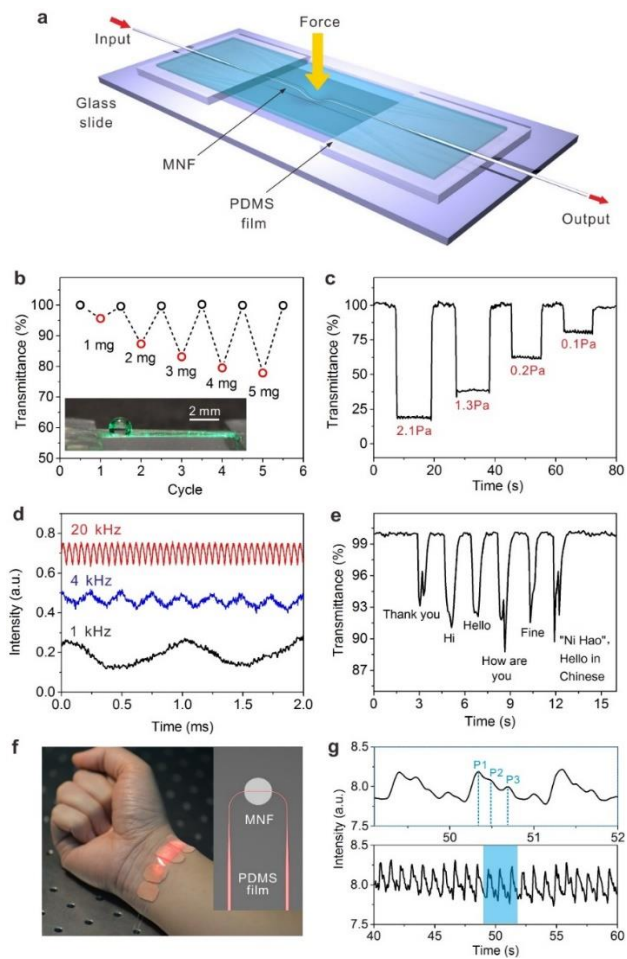

**Figure 3. Characterization of suspended and wearable O-skins. a,** Schematic of testing suspended O-skin. **b,** Response of a suspended O-skin to water droplets with different weights. The inset shows an optical micrograph of the O-skin with a water droplet atop. **c,** Response of a suspended O-skin to pressure of 2.1, 1.3, 0.2 and 0.1 Pa, respectively. **d,** Temporary response of a suspended O-skin to forced oscillation frequencies of 1, 4 and 20 kHz, respectively. **e,** Response of a suspended O-skin to acoustic vibrations from human voice. **f,** Photograph showing a skin-attachable O-skin directly above the artery of the wrist. The inset shows a schematic of the wearable O-skin. **g,** Measurement of wrist pulse under normal-condition (66 beats per minute).

Moreover, with pressure lower than 0.2 Pa, the O-skin offers a sensitivity of as high as 1870 KPa$^{-1}$, which is much more sensitive than that of high-performance electronic skin sensors (e.g., 0.55–192 KPa$^{-1}$)[22-24]. To investigate the temporary response, an O-skin (80-μm thickness, 1.2-μm MNF diameter) is used to measure mechanical vibration. The clear distinguishability up to 20 kHz (Fig. 3d), corresponding to a response time of ~10 μs, is more than three orders of magnitude faster than fast-response electronic-skin sensors (10-30 ms)[25-27]. In principle, without parasitic electrical effects, an O-skin should be able to respond even more rapidly, when a higher-frequency vibration test platform is available. Not surprisingly, the O-skin is highly sensitive to voice that propagates as acoustic waves in air (Fig. 3e), offering a possibility to resolve weak and rapid fluctuation of air pressure in real-time. Also, when being attached to a hand wrist (Fig. 3f), an O-skin can readily read out wrist pulse with high resolution (Fig. 3g), the two distinguishable peaks (P1 and P3) and a late systolic augmentation shoulder (P2), agree very well with the noninvasive high-fidelity recording of the radial artery pressure wave[28].

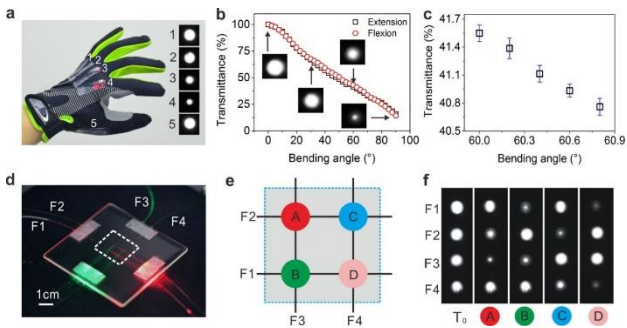

**Figure 4. Optical data gloves and O-skins with perpendicularly intersected 2×2 MNF arrays. a,** Photograph showing a five-sensor data glove integrated with 5 O-skins. **b,** Bending-angle-dependent output of a typical O-skin. **c,** Close-up view of the O-skin output within bending angles of 60.0-60.8°. **d,** Photograph showing an O-skin consisted by a perpendicularly intersected 2×2 MNF array. **e,** Schematic of the sensing areas for tactile sensing. **f,** Logic outputs of 4-input/output-port O-skin.

## 2.4 Optical data gloves

Individual O-skins can also be employed in multi-channel sensing. Figure 4a shows a five-sensor optical data glove for monitoring the flexion and extension of the metacarpophalangeal (MCP) joints of individual fingers, presenting a monotonic and approximately linear dependence of individual O-skin output on the bending angle (Fig. 4b) with an angular resolution of better than 0.2 °(Fig. 4c), which is much higher than that of the data gloves constructed from standard fiber-optic sensors (e.g., 1.8 °)[29] and stretchable conductive fiber strain sensors (e.g., 1.5 °)[30].

## 2.5 O-skins with MNF arrays

With negligible crosstalk (Fig. 1f), multiple MNFs can also be weaved inside a single O-skin for spatially resolved 2-dimensional tactile sensing. Figure 4d shows an O-skin (200-μm-thick) embedded with perpendicularly intersected 2×2 MNF arrays (15-mm separation between neighboring MNFs). When pressure is applied on the net-joint areas from node A to node D successively (Fig. 4e), logic readouts can be obtained (Fig. 4f): with zero external pressure ($T_0$), the four outputs read high; when the pressure is exerted on a certain point (e.g., node A), the two channels that cross the point (e.g., channels F2 and F3 at node A) give low output due to pressure-induced losses, while the others remain to be high. For spatially denser pressure sensing, the logic functionality of O-skin with smaller MNF separation (e.g., 3 mm) has also been realized. Finally, it is worth mentioning that, the power of the probe light (coupled from an LED) in the above-mentioned O-skin is about 400 nW, which can be further reduced through optimization, offering opportunities for ultralow-power operation.

## 3 Conclusion

We have demonstrated an O-skin that greatly surpasses conventional wearable sensors in sensitivity, response time and EMI immunity. In principle, the sensitivity of the O-skin can still be significantly improved when the refractive index of the host film approaches that of the MNF, and the response time is only limited by the mechanical properties of the hybrid MNF-PDMS structure. Meanwhile, the minimized optical crosstalk points toward the possibility of realizing multifunctional O-skin based on high-density optical circuitries, and the simple hybrid structures allow excellent versatility and large-scale fabrication. These initial results may pave the way towards future wearable optical devices for applications including human-machine interfaces, health monitoring and artificial intelligence.

## Acknowledgments

We thank the State Key Laboratory of Modern Optical Instrumentation and all members of the laboratory for support. This work was supported by the National Key Research and Development Program of China (2016YFB1001300), the National Natural Science Foundation of China (No. 11527901) and the Fundamental Research Funds for the Central Universities.